\documentclass[aps,twocolumn,prl,preprintnumbers,amsmath,amssymb,superscriptaddress]{revtex4-1}

\usepackage{graphicx}
\usepackage{bm}
\usepackage{hyperref}
\usepackage{xcolor}
\usepackage{IEEEtrantools}

\usepackage{float}

\restylefloat{table}

\begin{document}

\title{Terahertz-Induced Tunnel Ionization Drives Coherent \\Raman-Active Phonon in Bismuth}


\author{Bing Cheng}\email{chengbing986@gmail.com}
\affiliation{\mbox{Stanford Institute for Materials and Energy Sciences, SLAC National Accelerator Laboratory, Menlo Park, CA 94025, USA}}

\author{Patrick L. Kramer}
\affiliation{Laser Science and Technology, SLAC National Accelerator Laboratory, Menlo Park, CA 94025, USA}

\author{Mariano Trigo}
\affiliation{\mbox{Stanford Institute for Materials and Energy Sciences, SLAC National Accelerator Laboratory, Menlo Park, CA 94025, USA}}
\affiliation{ PULSE Institute, SLAC National Accelerator Laboratory, Menlo Park, CA, 94025, USA}

\author{Mengkun Liu}
\affiliation{Department of Physics and Astronomy, Stony Brook University, Stony Brook, New York 11794, USA}

\author{Ctirad Uher}
\affiliation{Department of Physics, University of Michigan, Ann Arbor, MI 48109, USA}

\author{David A. Reis}\email{dreis@stanford.edu}
\affiliation{\mbox{Stanford Institute for Materials and Energy Sciences, SLAC National Accelerator Laboratory, Menlo Park, CA 94025, USA}}
\affiliation{ PULSE Institute, SLAC National Accelerator Laboratory, Menlo Park, CA, 94025, USA}

\author{Zhi-Xun Shen}
\affiliation{\mbox{Stanford Institute for Materials and Energy Sciences, SLAC National Accelerator Laboratory, Menlo Park, CA 94025, USA}}
\affiliation{Geballe Laboratory for Advanced Materials, Stanford University, Stanford, CA 94305, USA}
\affiliation{Departments of Physics and Applied Physics Stanford University, Stanford, California 94305, USA}

\author{Jonathan A. Sobota}\email{sobota@stanford.edu}
\affiliation{\mbox{Stanford Institute for Materials and Energy Sciences, SLAC National Accelerator Laboratory, Menlo Park, CA 94025, USA}}

\author{Matthias. C. Hoffmann}\email{hoffmann@slac.stanford.edu}
\affiliation{Laser Science and Technology, SLAC National Accelerator Laboratory, Menlo Park, CA 94025, USA}
\date{\today}

\date{\today}

\begin{abstract}

Driving coherent lattice motion with THz pulses has emerged as a novel pathway for achieving dynamic stabilization of exotic phases that are inaccessible in equilibrium quantum materials. In this work, we present a previously unexplored mechanism for THz excitation of Raman-active phonons. We show that intense THz pulses centered at 1 THz can excite the Raman-active $A_{1g}$ phonon mode at 2.9 THz in a bismuth film. We rule out the possibilities of the phonon being excited through conventional anharmonic coupling to other modes or via a THz sum frequency process. Instead, we demonstrate that the THz-driven tunnel ionization provides a plausible means of creating a displacive driving force to initiate the phonon oscillations. Our work highlights a new mechanism for exciting coherent phonons, offering potential for dynamic control over the electronic and structural properties of semimetals and narrow-band semiconductors on ultrafast timescales.

\end{abstract}

\maketitle

\setlength{\parskip}{0.1em}

The use of ultrafast laser pulses to excite and manipulate coherent phonons presents a promising approach for dynamically controlling the electronic and structural properties of quantum materials \cite{bismuth_phonon_1990,Sokolowski-Tinten2003,Fritz2007,RMP_nonlinear_photonics}. As a result, the exploration of phonon driving mechanisms across different materials has becomes an important research subject. For the excitation of Raman-active phonons via femtosecond laser pulses at visible or near-infrared wavelengths, two primary excitation mechanisms, impulsive stimulated Raman scattering (ISRS)\cite{ISRS_NEW1,ISRS_1996} and displacive excitation of coherent phonons (DECP)\cite{DECP_1992,DECP_NEW1}, have been extensively discussed in the literature. The excitation of coherent phonons via ISRS is described by the equation in the single impulse limit: $\frac{d^2Q}{dt^2}+\Omega^2Q =\frac{1}{2}\left( \frac{\partial\chi}{\partial Q}\right)|E_0(\mathbf{r},t)|^2$. Here $Q$ is the amplitude of the Raman-active phonon mode, $\Omega$ is the phonon frequency, $E_0$ is the driving electric field of the laser pulse and $\chi$ is the electronic susceptibility. This equation indicates that the lattice motion of the Raman phonon is controlled by the Fourier components of the driving force at frequency $\Omega$. Efficient excitation is only possible when the laser pulse duration is short compared with $\Omega^{-1}$ and there is a sufficiently large coupling of the phonon coordinate $Q$ with the susceptibility $\chi$. On the other hand, in the DECP picture, the driving force of coherent phonon comes from the laser-induced sudden change of the carrier density or the effective electronic temperature\cite{DECP_1992}. If this laser-induced sudden change is fast enough, the ions will be driven to a new quasi-equilibrium position in a time scale shorter than $\Omega^{-1}$. Such an ultrafast change of ion position generates a restoring force for the coherent atomic motion of the Raman phonon.

Despite the  well-established ISRS and DECP mechanisms, the recent observations of THz driving Raman-active phonons in solids are challenging these simple pictures \cite{THz_SHG_phonon_2016,THz_SHG_phonon_2017,THz_phonon_dimond_2017,THz_phonon_BiSe_2018,THz_phonon_CdWO4_2019,THz_phonon_BiSe_2020,THz_pump_phonon_2022terahertz,THzsum_2024}. The oscillation cycle of THz pulses generated by the tilted pulse front method is approximately 1 ps \cite{Hebling:02_THz}. Based on the assumptions of ISRS and DECP, it is somewhat counterintuitive to see such ``slow'' THz pulses drive Raman-active phonons above 1 THz. However, in some metallic materials like doped Bi$_2$Se$_3$, intense THz pulses have been found to drive Raman-active phonon modes at 2.05 THz and 4.05 THz \cite{THz_phonon_BiSe_2018,THz_phonon_BiSe_2020}. Competing mechanisms have been proposed to understand these unusual observations. The first proposal involves an anharmonic phonon coupling model. In this model, the intense THz pulses first excite an infrared-active zone center phonon mode \cite{THz_phonon_BiSe_2018,THz_phonon_BiSe_2020}. The anharmonic interactions between the phonons allow the THz-driven oscillations of the infrared-active mode to serve as an effective driving force to launch Raman-active phonon modes. Another mechanism under consideration is THz sum frequency generation. In this scenario, the excitation of Raman-active phonon modes occurs directly through the absorption of two THz photons via a Raman-type process that does not require the intermediary action of other phonons \cite{THz_phonon_dimond_2017,THz_phonon_CdWO4_2019}. Both this model and the anharmonic phonon coupling model have been effective in elucidating various nonlinear phononic and photonic phenomena observed in materials \cite{Cavalleri_2011,THz_phonon_CdWO4_2019}.

In this work, we investigate a novel but previously-unexplored mechanism for excitation of Raman-active phonons in semimetallic bismuth (Bi) thin films though THz pulses. Bismuth is a prototypic material that has been extensively investigated by various ultrafast and nonlinear spectroscopies \cite{Bi_phonon_1996,Sokolowski-Tinten2003,Bi_phonon_2006,Fritz2007,Bi_phonon_new,Bi_film_resource_2013,MIR_bismuth,2DTHz_bismuth}. Previous optical pump-probe experiments on bismuth identified two Raman-active phonons: an E$_g$ mode at 2.1 THz and an A$_{1g}$ mode at 2.9 THz \cite{Bi_phonon_1996,Bi_phonon_2006,Bi_phonon_new,Bi_film_resource_2013}. Using THz-pump optical-probe spectroscopy, we observed long-lived coherent oscillations of the A$_{1g}$ mode at 2.9 THz in reflectivity at 800 nm. Our analysis suggests that the excitation of the A$_{1g}$ mode is likely due to THz-induced tunnel ionization, which rapidly increases carrier density and consequently triggers the phonon.

\setlength{\belowcaptionskip}{-0.5cm} 

\begin{figure}[t]
\includegraphics[width=0.33\textwidth]{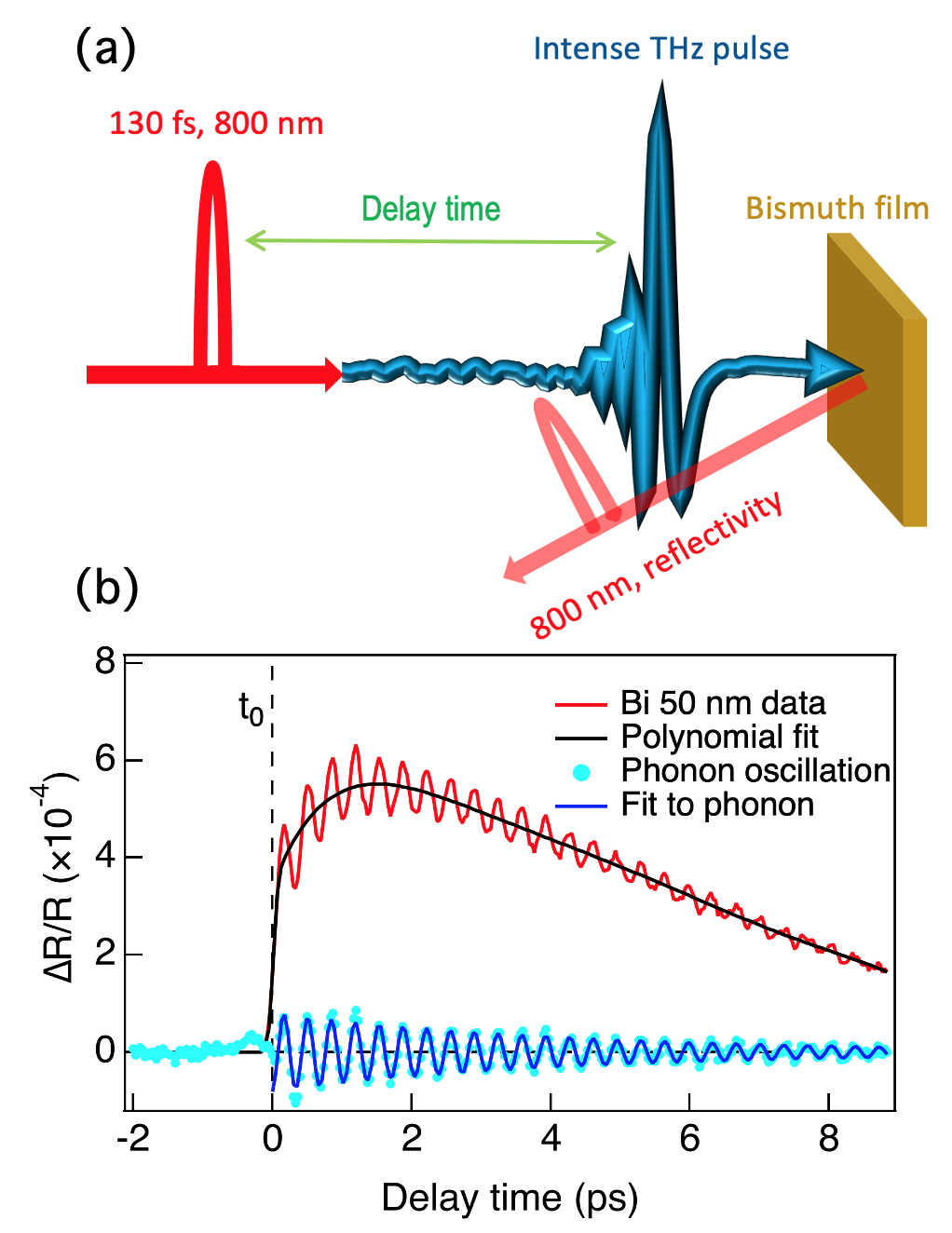}
\caption{\label{fig:Bi50nm_THz_pump_long} (a) Schematic of THz pump optical probe spectroscopy. (b) THz-induced reflectivity change $\Delta R/R$ of a 50 nm thick bismuth film as a function of pump-probe delay time at 400 kV/cm. We use a cosine function to fit the phonon oscillation isolated by a polynomial fit. The initial phase of the phonon oscillation is found to be (0.035 $\pm$ 0.009)$\pi$.}
\label{Fig1}
\end{figure}

\setlength{\belowcaptionskip}{-0.5cm} 

\begin{figure}[t]
\includegraphics[width=0.37\textwidth]{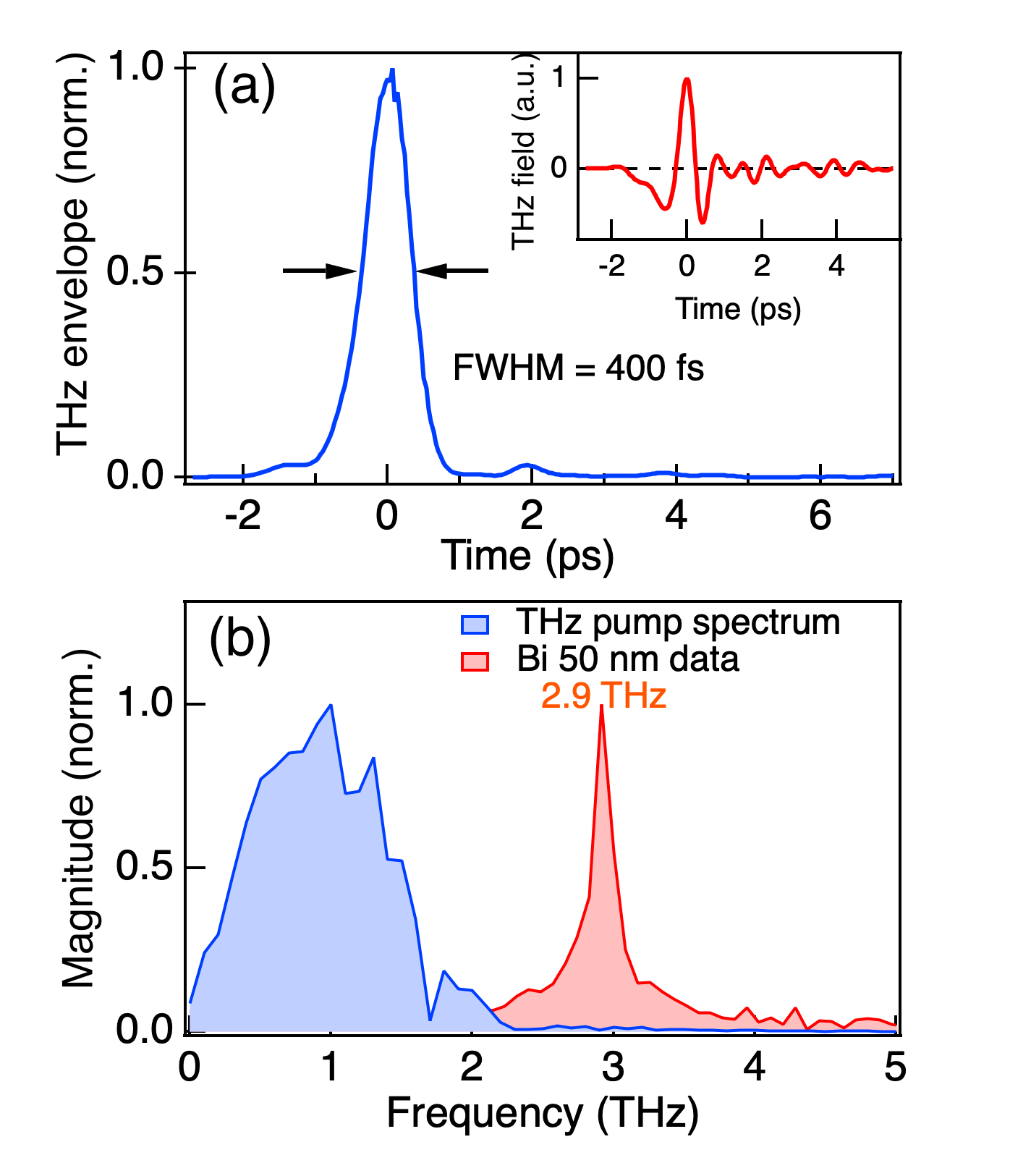}
\caption{\label{fig:Bi50nm_THz_pump_long}  (a) The THz pulse width is defined by the full width at half maximum (FWHM) of the square of the Hilbert transform of the THz field. The inset shows THz field measured by electro-optic sampling.  (b) Spectral magnitude of phonon oscillations (red) and THz pulse (blue) obtained by Fourier transform.}
\label{Fig2}
\end{figure}

The Bi films of 50 nm thickness were deposited in the molecular beam epitaxy (MBE) system from a Knudsen cell on 1 $\times$ 1 cm$^2$ sapphire substrates (c-axis 001 cut, $\sim$0.5 mm thick). The films were single crystals with the c (trigonal) axis perpendicular to the surface as determined by x-ray diffraction. The sample used in this study is similar to the ones used in Ref. \cite{Bi_film_resource_2013}. We used single-cycle THz pulses up to 500 kV/cm generated from optical rectification in LiNbO$_3$ to excite Bi film. The polarization of THz pulses is perpendicular to the trigonal axis. The pump-induced reflectivity change was probed by a 130 fs, 800 nm pulse.  The schematic of our THz pump optical probe setup is shown in Figs. \ref{Fig1}(a). More details are provided in Supplementary Material \cite{SI}. All measurements in this work were performed at room temperature. Figure \ref{Fig1}(b) shows the reflectivity change, $\Delta R/R$, as a function of delay time at a THz field strength of 400 kV/cm. The polarization of the probe pulse was perpendicular to that of THz pulse. Following a fast initial rise, clear coherent phonon oscillations with a period of 340 fs are observed against a slowly decaying electronic background. These oscillations correspond to a mode frequency of 2.9 THz (Figs. \ref{Fig2}(b)), which aligns with the well-known A$_{1g}$ Raman-active mode in bismuth. At the frequency corresponding to the A$_{1g}$ phonon, the THz field strength is about 1\% of its peak value.

\begin{figure*}[t]
\centering
\includegraphics[width=0.9\textwidth]{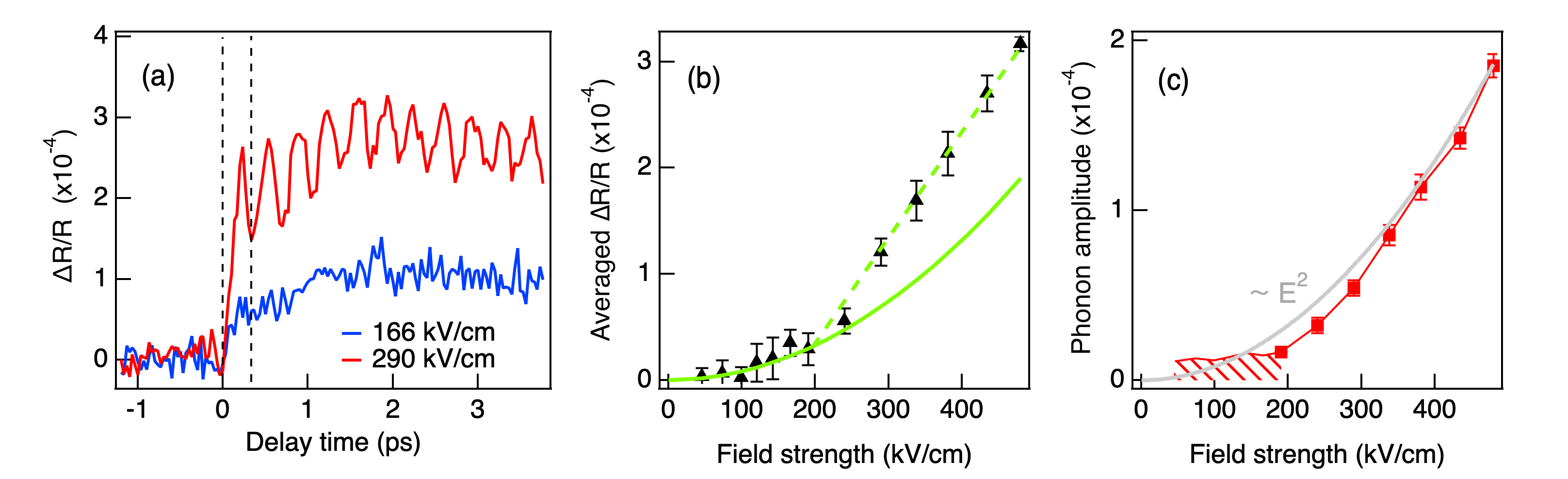}
\caption{\label{fig:Bi50nm_THz_pump_long} (a) Time-resolved reflectivity change $\Delta R/R$ for two representative THz pump field strength of 166 and 290 kV/cm. (b) Average reflectivity change in the time interval of 0 to 0.4 ps labelled by the black dashed lines as a function of THz field strength. (c) Phonon oscillation amplitude as a function of THz field strength. It deviates from the quadratic field dependence (gray curve). Below 200 kV/cm, no phonon oscillations are observed (hatched area).}
\label{Fig3}
\end{figure*}

Figure \ref{Fig2}(a) displays the THz intensity, defined as the square of the field envelope. This envelope was derived from an electro-optical measurement of the THz field at the sample location (inset) using Hilbert transformation. Notably, the full width at half maximum (FWHM) of the THz intensity is 400 fs, which exceeds the oscillation period of the bismuth A$_{1g}$ phonon (340 fs). As a result, a simple impulsive or displacive model cannot explain the observed excitation at 2.9 THz. We need to explore models beyond conventional mechanisms to understand how this A$_{1g}$ phonon is driven by THz pulses.

In order to elucidate the excitation mechanism, we carried out a THz field strength dependent study. Figure \ref{Fig3}(a) shows representative time traces of reflectivity change $\Delta R/R$ at two different pump field strengths. At a field strength of 290 kV/cm, a sharp rise in reflectivity is observed and the coherent phonon oscillations are clearly visible. In contrast, at 166 kV/cm and within our detection limit, the rapid rise near time zero is extremely weak. The reflectivity from 0 to 0.4 ps is dominated by a slower rise process, and coherent phonon oscillations are no longer observed under this lower pump field condition. These contrasting responses indicate the existence of a threshold effect influencing the phonon excitation. To quantify this threshold behavior, we calculated both the average reflectivity change in the 0 -- 0.4 ps window after excitation and the amplitude of the coherent phonon oscillation at various THz field strengths. The complete field strength dependent $\Delta R/R$ data and the descriptions of our methods can be found in the Supplementary Material \cite{SI}. Figure \ref{Fig3}(b) and \ref{Fig3}(c) show the average signals and the extracted phonon oscillation amplitude as a function of THz field strength up to 480 kV/cm, respectively. Both plots exhibit a clear threshold behavior. Below $\sim$ 200 kV/cm, the average $\Delta R/R$ signal scales with the square of the THz field strength, and the phonon oscillation amplitude is negligible. As the field strength is beyond 200 kV/cm, the average signal increases at a much faster rate and the phonon oscillation amplitude starts to become significant. This threshold behavior directly correlates the fast rising edge in the reflectivity data with the onset of coherent phonon oscillations.

Before delving into the details of the excitation mechanism, it is essential to evaluate if current prevailing models can explain our results. THz driving of Raman-active phonons has been observed in various materials and interpreted using either an anharmonic coupling model or a THz sum frequency model \cite{THz_SHG_phonon_2016,THz_SHG_phonon_2017,THz_phonon_dimond_2017,THz_phonon_BiSe_2018,THz_phonon_CdWO4_2019,THz_phonon_BiSe_2020,THz_pump_phonon_2022terahertz,THzsum_2024}. Based on our data, we can swiftly exclude the conventional anharmonic coupling model that relies on the presence of zone-center infrared-active optical phonon modes \cite{THz_phonon_BiSe_2018,THz_phonon_BiSe_2020}. Bismuth has a rhombohedral A7 lattice structure, with two Bi atoms in each unit cell. The phonon dispersion of the bismuth lattice does not include any zone-center infrared-active optical phonon modes in the THz regime \cite{Bi_cell_phonon_2007,Bismuth_phonon_cal}. On the other hand, our simulation of phonon dynamics with THz sum frequency model does not reproduce the sharp rise edge on $\Delta R/R$. Details of the simulation are presented in the Supplemental Material \cite{SI}. In addition, the model predicts that the driven phonon amplitude scales quadratically with the THz field strength. However, as shown in Figs. \ref{Fig3}(c), the bismuth phonon amplitude exhibits a threshold followed by a quasi-linear dependence on the THz field, deviating from the quadratic field behavior. These observations suggest that the THz sum frequency model is unlikely to explain our results.

\begin{figure*}[t]
\centering
\includegraphics[width=0.8\textwidth]{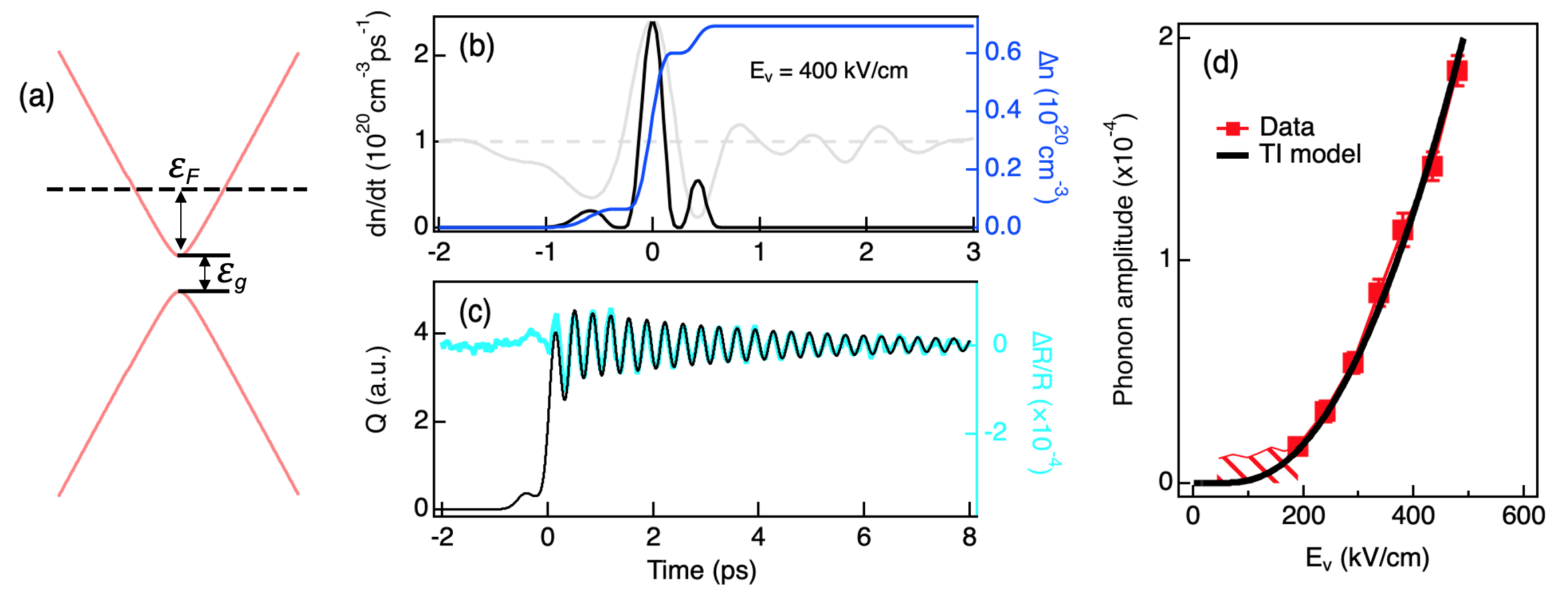}
\caption{\label{fig:Bi50nm_THz_pump_long} (a) Schematic for band structure of bismuth near the L points. $\mathcal{E}_g$ is the band gap and $\mathcal{E}_F$ is the Fermi energy. (b) The temporal profile of the transient tunneling rate $dn/dt$ and the transient injected carrier density $\Delta n$ at a field strength of 400 kV/cm. For comparison, the waveform of THz pulse (gray curve) in time domain is also plotted. (c) Phonon dynamics at 400 kV/cm simulated by the tunnel ionization (TI) model. The experimental phonon oscillation at the same field strength is also plotted with offset for comparison. (d) Simulated phonon amplitude as a function of THz field strength plotted together with the phonon amplitude determined by our experiments.  }
\label{Fig4}
\end{figure*}

Elemental bismuth is a semimetal with low carrier density and small band gap \cite{Bismuth_infarred_2010}.  Figure \ref{Fig4}(a) presents its schematic band structure near the L points, where the band gap $\mathcal{E}_g$ is approximatively 15 meV \cite{Bismuth_parameters}. Previous studies have demonstrated that the presence of strong electric field can produce additional carriers via tunnel ionization in semiconductors \cite{keldysh1958behavior,keldysh1964ionization}. As illustrated by Figs. \ref{Fig4}(a), because of Pauli blocking, electrons in the valence bands need to overcome a minimum energy barrier in the order of $\mathcal{E}_g + \mathcal{E}_F$ to transition into the conduction bands---a threshold significantly higher than the energy of THz photons. To simplify the analysis of tunnel ionization in bismuth, we can treat it as a semiconductor, with electrons in the valence bands effectively bound by a potential barrier $\mathcal{E}_g^{\prime}$ $\sim$ $\mathcal{E}_g + \mathcal{E}_F$. The presence of an intense THz electric field can significantly distort this potential barrier, enabling tunnel ionization where electrons can tunnel through the potential barrier \cite{keldysh1958behavior,keldysh1964ionization}. As a result, a rapid increase of the carrier density may be induced, which can be a driving force of the A$_{1g}$ phonon \cite{DECP_1992}.

To assess if THz-driven tunnel ionization can yield a sufficiently rapid rise in the injected carrier density $\Delta n$ to initiate the $A_{1g}$ phonon, we use the following formula, which is valid in the effective mass approximation, to simulate the carrier tunneling rate $dn/dt$ of valance electrons transitioning to the conduction bands \cite{keldysh1958behavior}.

 \vspace{-0.3cm}

\begin{equation}
\frac{dn}{dt}=2\pi (\frac{eE_{in}}{2\pi \hbar})^2(\frac{m_1m_2m_3}{m_{||}^2\mathcal{E}_g^{\prime}})^{\frac{1}{2}}e^{-\frac{\pi}{2e\hbar E_{in}}\sqrt{m_{||}(\mathcal{E}_g^{\prime})^3}}
\label{TI}
\end{equation}

\noindent Here $E_{in}$ is the THz electric field in bismuth film. $E_{in}$ is related to the THz electric field in vacuum, $E_v$, by a factor $\alpha$: $E_{in} = \alpha E_v$. At room temperature, $\alpha$ is estimated to be $\sim$ 0.25. $\mathcal{E}_g^{\prime}$ is the potential barrier and approximately equates $\mathcal{E}_g + \mathcal{E}_F$. $m_1$, $m_2$ and $m_3$ are determined by $m_i^{-1}=m_{ic}^{-1}+m_{iv}^{-1}$ ($i$ = 1, 2, 3), where 1, 2, and 3 refer to the binary, bisectrix, and trigonal axes, respectively. $m_{ic}$ and $m_{iv}$ are the effective masses of the bottom of conduction bands and the top of valance bands, respectively. $m_{||}$ is determined by $m_{||}^{-1}$ = cos$^2$$\theta$/$m_1$ + sin$^2$$\theta$/$m_2$. Here $\theta$ is the angle between THz polarization and the principal axis. We take the numerical values of $m_1$, $m_2$, $m_3$, $\mathcal{E}_g$ and $\mathcal{E}_F$ from Ref. \cite{Bismuth_parameters}. For capturing the threshold behavior of the phonon excitation, we rescale these parameters to match the threshold field strength $\sim$ 200 kV/cm. More details of the simulation are presented in the Supplemental Material \cite{SI}. Figure \ref{Fig4}(b) shows the simulated temporal profile of $dn/dt$ at $E_v$ = 400 kV/cm. It is evident that the temporal profile of $dn/dt$ is much sharper than the THz pulse. By integrating $dn/dt$ over time, we obtain the temporal profile of injected carrier density $\Delta n$. As shown in Figs. \ref{Fig4}(b), $\Delta n$ undergoes a sharp rise near time zero, on a timescale of $\sim$ 200 fs. This rapid increase should be fast enough to drive the $A_{1g}$ mode in bismuth \cite{DECP_1992}.

We now extend our analysis to the phonon dynamics simulation. Specifically, when incorporating the tunnel ionization model, the phonon motion equation takes the following form \cite{DECP_1992}:

 \vspace{-0.3cm}

\begin{equation}
\frac{d^2Q}{dt^2}+\gamma\frac{dQ}{dt}+\omega^2_{0}Q=k\Delta n(t)
\label{anharmonic}
\end{equation}

Here $Q$ represents the driven phonon displacement and $\gamma$ is the damping coefficient. $\omega_0$ is the angular frequency of the $A_{1g}$ mode. The constant $k$ linearly relates the change in carrier density to the electrostatic force acting on the $A_{1g}$ mode. Additional details about the simulation are provided in the Supplemental Material \cite{SI}. Figure \ref{Fig4}(c,d) show that our simulation successfully reproduces several key experimental observations. First, the phonon displacement $Q$ exhibits a rapid rise near time zero, which corresponds to the steep onset of the THz-induced reflectivity change $\Delta R/R$. Second, the simulated phonon oscillations agree well with the experimentally observed oscillations, as highlighted in Figs. \ref{Fig4}(c). Third, the displacement $Q$ demonstrates a pronounced threshold behavior with respect to the THz field strength. The complete results are provided in the Supplemental Material \cite{SI}. To highlight this threshold behavior, we extracted the phonon oscillation amplitude from $Q$ and plotted it as a function of THz field strength in Figs. \ref{Fig4}(d). The resulting field dependence clearly exhibits a threshold behavior and deviates from the $E^2$ trend. Notably, the simulated phonon amplitude completely reproduces the THz field strength dependence of experimental data, confirming that our tunneling-based model accurately captures the essential physics. Finally, we note that our current simulation does not account for carrier relaxation effects, which would be necessary to describe the extended carrier relaxation of $\Delta R/R$ at longer times. Here, we focus on the phonon excitation governed by carrier tunneling near time zero. Even within this scope, our tunnel ionization model effectively reproduces the key features of THz-driven phonon dynamics in bismuth.

The observation of THz-driven Raman-active phonons via tunnel ionization represents a significant advance in understanding coherent phonon driving mechanisms. This discovery establishes the first direct experimental evidence that tunnel ionization can serve as a viable route to coherent phonon excitation and bridges two previously disconnected areas—strong-field ionization and coherent phonon dynamics. Crucially, in this new pathway, the driven phonon amplitude exhibits an exponential dependence on THz field strength $E$, i.e., $Q \propto E^2e^{-a/E}$ ($a$ is a constant), in stark contrast to established excitation pathways that mainly  follow $Q \propto E^2$. Moreover, previous studies have demonstrated that THz pulses can effectively drive IR-active phonons by coupling to their dipole moments \cite{IRphonon_THz_2020,KTO_SHG}, as well as Raman-active phonons through nonlinear phononics \cite{mitrano2016possible,THz_phonon_BiSe_2018}. These capabilities open up opportunities to explore light-induced phase transitions via lattice control in quantum materials \cite{mitrano2016possible,Li1079,Nova1075,KTO_SHG}. Our discovery provides a new route for dynamically tuning the electronic and structural properties of semimetals and narrow-band semiconductors where THz-driven tunnel ionization can play important roles. Compared with conventional excitation methods using near-infrared or visible light, THz-driven tunnel ionization avoids high-energy interband transitions, thereby suppressing unwanted thermal dissipation and heating. This advantage is crucial for optically controlling fragile low-temperature quantum phases. A particularly promising example lies in the twisted moir\'{e} two-dimensional materials, which have recently garnered intense interest for their rich correlated and topological phenomena at low temperatures. In these systems, moir\'{e} superlattices typically generate narrow band gaps on the order of tens of meV\cite{Andrei2020,Mak2022}, making them susceptible to THz-driven tunnel ionization. Such sensitivity offers unique opportunities for real-time manipulation of correlated and topological states in moir\'{e} quantum materials via selective lattice control. In addition to expanding the mechanisms for coherent phonon excitation and the potentials for dynamical control of materials, our work further features the potential roles of carrier tunnelling effect itself in semimetals and correlated materials under strong THz pump. This is particularly true for insulator-metal transitions where this becomes more important as the gap closing makes tunnel ionization and ultimately single-photon ionization possible \cite{MIT}. Thus, the concept of driving materials in their electronic ground state may be deeply flawed or at least subject to  limitations. When interpretating THz-driven phenomena in these materials, it may be crucial to consider the roles of charge tunnel ionization.

In summary, we studied THz excitation of the $A_{1g}$ phonon mode in a bismuth thin film. We showed that this phonon is excited through a carrier tunnel ionization process rather than an anharmonic phonon coupling or a THz sum frequency mechanism. We demonstrated that the THz-driven tunnel ionization provides an effective means of sharpening the rising edge of the tunneling carrier density, creating a displacive driving force triggering the phonon oscillations. This provides an alternative approach to selective control of coherent phonons in solids. Our findings can be easily extended to other semimetals or semiconductors and illustrate the importance of carrier mediated effects like tunnel ionization in experiments using high-field THz pulses. 

We would like to thank Liangbo Liang for helpful discussion. Use of the Linac Coherent Light Source, SLAC National Accelerator Laboratory, is supported by the US Department of Energy (DOE), Office of Science, Basic Energy Sciences, under contract no. DE-AC02-76SF00515. The work at SIMES (BC, DAR, JAS, MT and ZXS) is supported by the U.S. Department of Energy (DOE), Office of Science, Basic Energy Sciences, Materials Sciences and Engineering Division under Contract No. DE-AC02-76SF00515; the work at LCLS (PK and M.C.H) is supported by U.S. Department of Energy (DOE), Office of Science, Basic Energy Sciences, under award no. 2018-SLAC-100499. The work at Stony Brook University (M.K.L.) is supported by the NSF Faculty Early Career Development Program under Grant No. DMR-2045425, a QuantEmX travel grant from ICAM, and the Gordon and Betty Moore Foundation through Grant GBMF9616 to M.K.Liu.

\nocite{Simulation_THzfield_strength_2020,Bismuth_THzcond1,THzsum_theory111_2018}

 \bibliography{Quadratic}

 \newpage

\setcounter{figure}{0}
\setcounter{equation}{0}
\setcounter{section}{0}
\begin{widetext}

\maketitle

\section{S\lowercase{upporting} N\lowercase{ote} 1:  S\lowercase{ample growth and} TH\lowercase{z-pump optical-probe setup} }

\renewcommand{\thefigure}{S\arabic{figure}}

The Bi films of 50 nm thickness were deposited in the molecular beam epitaxy (MBE) system from a Knudsen cell on 1$\times$1 cm$^2$ sapphire substrates (c-axis 001 cut, $\sim$0.5 mm thick). The films were single crystal with the c (trigonal) axis perpendicular to the surface as determined by x-ray diffraction (XRD). The growth rate of Bi films was kept at 1.9$-$2.0 $\AA$/s. The growth conditions were in-situ monitored by Reflection High Energy Electron Diffraction (RHEED) while the films were being deposited. The thickness of bismuth films was determined by both the in-situ crystal thickness monitor and by monitoring the CCD camera that filmed the film's growth. The film quality was ex-situ examined by XRD measurements utilizing a Rigaku Ultima IV x-ray diffractometer with Cu K$_\alpha$ radiation.

The schematic of our THz pump optical probe setup is depicted in Fig. \ref{Fig1_SI}. The single-cycle THz pulses were generated by optical rectification of 130 fs 800 nm laser pulses with 4 mJ energy using the tilted pulse front method in LiNbO$_3$. The THz pulses with energies up to 3 $\mu$J were focused onto the sample yielding a spot size of about 1.2 $\times$ 1.2 mm$^2$ and a fluence of $\approx$ 0.2 mJ/cm$^2$. The polarization of THz pulses is perpendicular to the trigonal axis of Bi film. The THz field strength at the focus was measured by electro optical (EO) sampling in a 100 $\mu$m thick 110-cut GaP crystal. The maximum field strength was $\sim$ 500 kV/cm. The waveform of THz pulse in time domain is displayed in main text Fig. 2. A pair of wire grid polarizers, controlled by a computerized rotation stage, was used to continuously attenuate the THz field. The linearity of this attenuation was verified by EO sampling. Weak probe pulses at 800 nm wavelength were directed onto the sample collinearly with the THz beam, and then reflected onto a photo diode by slightly tilting the sample. A part of the probe beam was split off before the sample and directed onto a second, identical, photo diode for balanced detection, minimizing laser noise. All measurements in this work were performed at room temperature.

\begin{figure}[h]
\includegraphics[clip,width=3.5in]{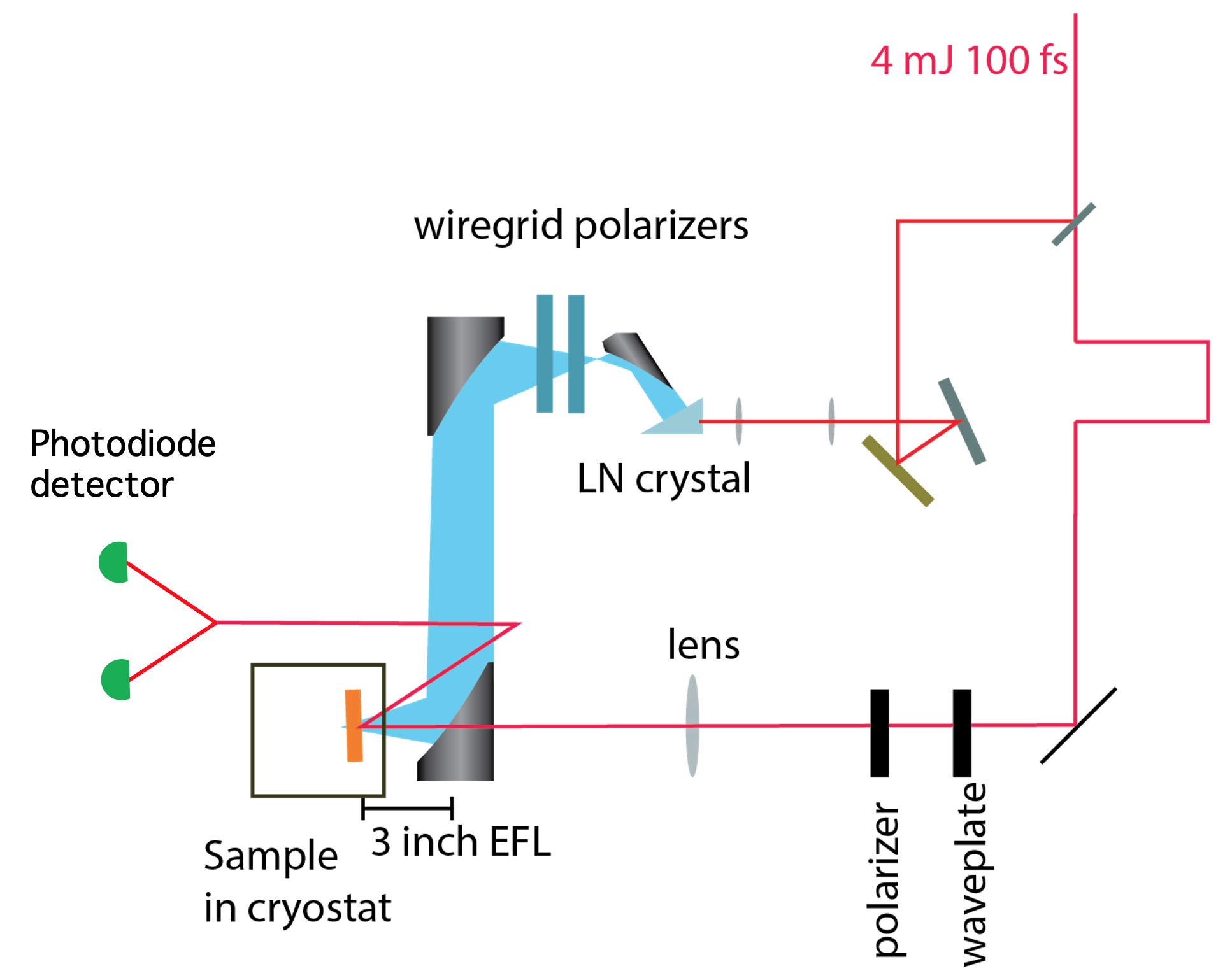}
\caption{(Color online) The schematic of our THz pump optical probe setup.}
\label{Fig1_SI}
\end{figure}

\clearpage

\section{S\lowercase{upporting} N\lowercase{ote} 2: F\lowercase{ull reflectivity data as a function of terahertz field strength}}

In Fig. \ref{Fig2_SI}, we show the reflectivity change $\Delta R/R$ of bismuth thin film as a function of delay time between THz pump and 800 nm probe pulses at various THz field strengths. One can see that below 200 kV/cm, the initial rise of $\Delta R/R$ at 0 to 0.4 ps is much slower. In the meanwhile, the coherent phonon oscillations are completely absent in the signals. In contrast, above 200 kV/cm, a sharp rise in reflectivity gradually develops and the coherent phonon oscillations are clearly visible. This full dataset is completely consistent with the excitation threshold behavior of the coherent phonon discussed in the main text.

\begin{figure}[h]
\includegraphics[clip,width=4in]{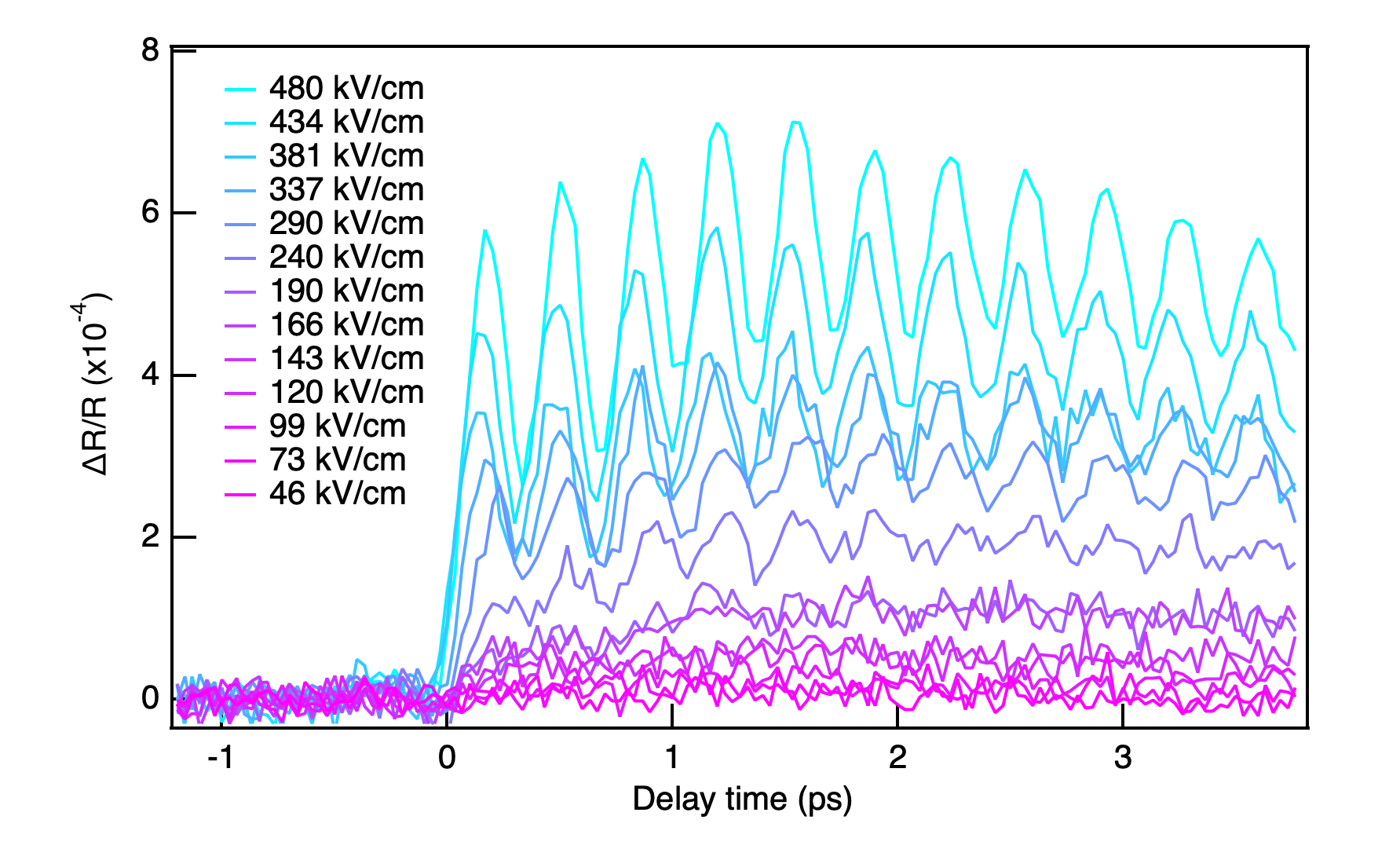}
\caption{(Color online) The reflectivity change $\Delta R/R$ of bismuth thin film as a function of delay time at various THz field strengths.}
\label{Fig2_SI}
\end{figure}

\clearpage

\section{S\lowercase{upporting} N\lowercase{ote} 3:  S\lowercase{ubstraction of electronic background} }

To extract the THz field strength dependent phonon amplitude, one needs to subtract the electronic background of the reflectivity change $\Delta R/R$ first. We used a polynomial function convolving with the step function to capture the electronic background. The equation is shown below.

\begin{equation}
\Delta R/R=a_0+(k_0+k_1t+k_2t^2+k_3t^3+k_4t^4)(1+erf(2\sqrt{ln2}(t-t_0)/w))/2
\label{anharmonic}
\end{equation}

\noindent Here $a_0$ is the constant offset. $k_0$, $k_1$, $k_2$, $k_3$, and $k_4$ are the coefficients of the polynomial. $t_0$ is the time zero. $erf$ is the error function to simulate the rise of $\Delta R/R$. $w$ is the rise width. Note that in this work, we are only interested in THz pump field strength dependence of the phonon amplitude. The way to simulate electronic background will not modify the pump field strength dependence of phonon oscillations. The polynomial fitting is the easiest way to capture the electronic background. We will not discuss the physics meaning of these fitting. Our aim is only to isolate the phonon oscillations. We present all data fitting in Fig. \ref{Fig3_SI}. One can see the polynomial fitting can reasonably capture the electronic background. All residual phonon oscillation data are present in Fig. \ref{Fig4_SI}.

\begin{figure}[htp]
\includegraphics[clip,width=5.4in]{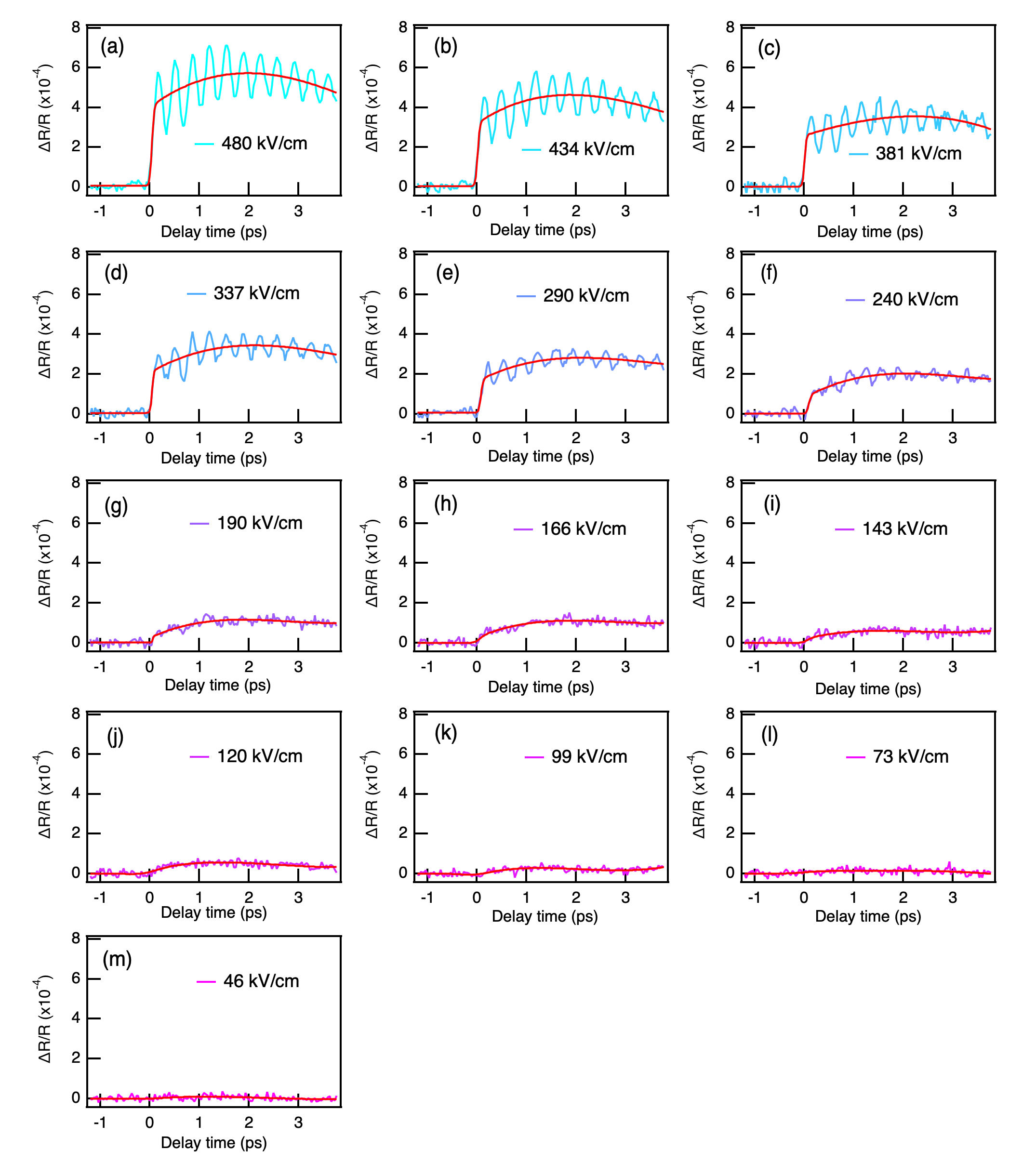}
\caption{(Color online) The fitting of the electronic background of $\Delta R/R$ at various THz field strengths.}
\label{Fig3_SI}
\end{figure}

\clearpage

\section{S\lowercase{upporting} N\lowercase{ote} 4:  F\lowercase{its of the phonon amplitude} }

To extract the THz field strength dependent phonon amplitude, we used a cosine function convolving with an exponential decaying function to fit the phonon oscillations. The equation is shown below.

\begin{equation}
\Delta R/R=Acos(2\pi f(t-t_0)+\phi_0)*exp(-(t-t_0)/\tau)
\label{anharmonic}
\end{equation}

\noindent Here $A$ is the phonon amplitude. $f$ is the phonon frequency. $t_0$ is the time zero. $\tau$ is the decay constant. $\phi_0$ is the initial phase of the phonon oscillations. We present all fittings in Fig. \ref{Fig4_SI}. The fittings reasonably capture all the phonon oscillations data above 200 kV/cm. Below 200 kV/cm, the phonon oscillations are not visible in the signals. We present the phonon oscillation amplitude as a function of THz field strength in the Fig. 3(c) of main text. Note that below 200 kV/cm, we do not observe phonon oscillations. Any fittings to these data become meaningfulness. We calculated the standard deviation of each curve in Fig. \ref{Fig4_SI}(g) and then plotted them as the hatched region in Fig. 3(c) of main text. This hatched region represents the upper limit on the phonon amplitude below 200 kV/cm.

\begin{figure}[h]
\includegraphics[clip,width=6.6in]{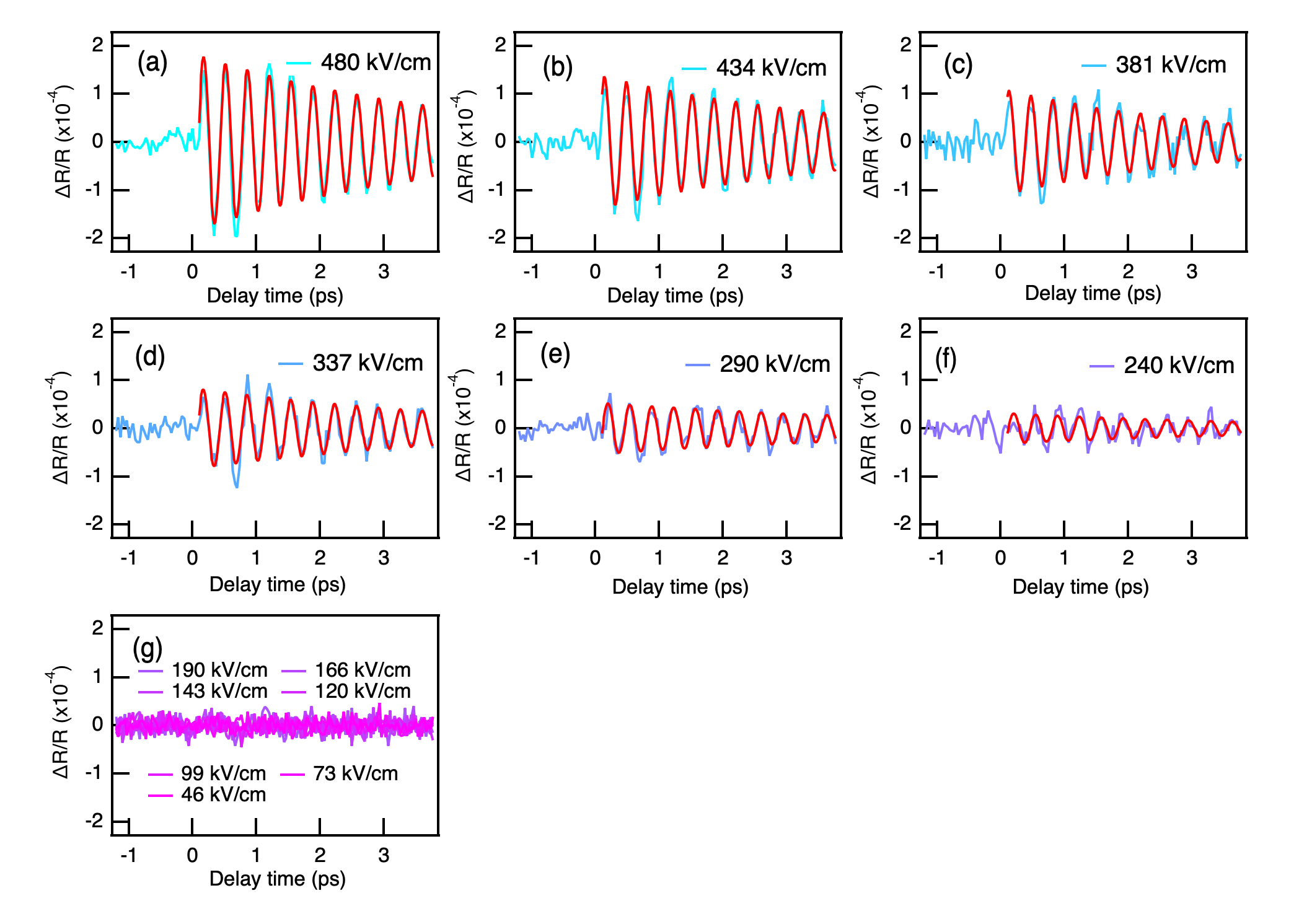}
\caption{(Color online) The fitting of the phonon oscillations at various THz field strengths.}
\label{Fig4_SI}
\end{figure}

\clearpage

\section{S\lowercase{upporting} N\lowercase{ote} 5:  E\lowercase{stimation of} TH\lowercase{z field strength inside the bismuth film} }

To simulate the tunneling rate $dn/dt$, we need to know the THz field strength $E_{in}$ inside the bismuth film. $E_{in}$ is related to the THz electric field in vacuum, $E_v$, by a factor $\alpha$: $E_{in} = \alpha E_v$. Here the coefficient $\alpha$ is determined by the following formula. \cite{Simulation_THzfield_strength_2020}

\begin{equation}
\alpha= \frac{1}{1-\frac{n_b-1}{n_b+1}\frac{n_b-n_s}{n_b+n_s}e^{2i\frac{\omega dn_b}{c}}}\frac{2}{n_b+2}(1+\frac{n_b-n_s}{n_b+n_s}e^{2i\frac{\omega dn_b}{c}})
\label{Note5}
\end{equation}

\noindent Here $n_b$ is the complex refractive index of the bismuth thin film. $n_s$ (3.3) is the refractive index of the sapphire substrate in THz regime. $d$ (50 nm) is the thickness of the bismuth film. To determine $\alpha$ at room temperature, we need to know the refractive index of the bismuth thin film $n_b$ at the room temperature. We used the terahertz conductivity data of bismuth thin film in Ref. \cite{Bismuth_THzcond1} to estimate $n_b$. Note that Ref. \cite{Bismuth_THzcond1} only provides terahertz conductivity data of bismuth film at 25 K. Bismuth is a semimetal so that its room temperature conductivity is usually larger than the conductivity at 25 K. For simplicity, we rescale a factor of 2 to the data at 25 K and use this rescaled data as room temperature terahertz conductivity. Then $n_b$ at room temperature is determined by $n_b=\sqrt {1+\frac{i\sigma (\omega)}{\epsilon_0 \omega}}$. Here $\epsilon_0$ is the permittivity of free space. $\sigma (\omega)$ is the optical conductivity. $\omega$ is the angular frequency. Fig. \ref{Fig4_SI}(b) presents the simulation of $\alpha$ at room temperature using $\sigma$ in Fig. \ref{Fig4_SI}(a). $\alpha$ shows very weak frequency dependence below 2 THz. We take $\alpha$ $\sim$ 0.25 to simulate the tunneling rate $dn/dt$ at room temperature.

\begin{figure}[h]
\includegraphics[clip,width=6.6in]{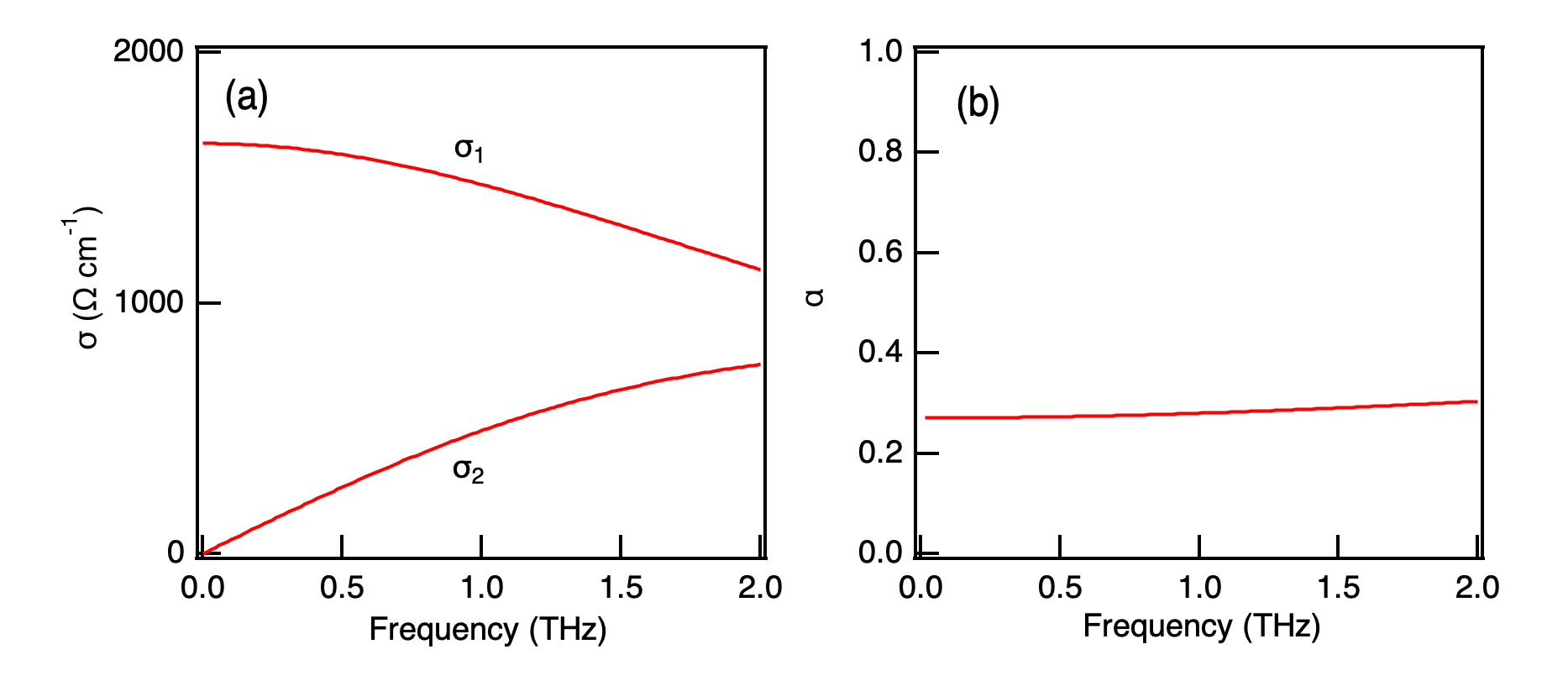}
\caption{(Color online) (a) Optical conductivity of bismuth thin film at room temperature estimated from the data presented in Ref. \cite{Bismuth_THzcond1}. (b) The simulation result of $\alpha$.}
\label{Fig5_SI}
\end{figure}

\clearpage

\section{S\lowercase{upporting} N\lowercase{ote} 6:  S\lowercase{imulation of} TH\lowercase{z field driving tunnel ionization in bismuth thin film} }

In the main text, we used below formula to simulate the tunneling rate $dn/dt$ induced by THz pulses in the bismuth film.\cite{keldysh1958behavior}

\begin{equation}
\frac{dn}{dt}=2\pi (\frac{eE_{in}}{2\pi \hbar})^2(\frac{m_1m_2m_3}{m_{||}^2\mathcal{E}_g^{\prime}})^{\frac{1}{2}}e^{-\frac{\pi}{2e\hbar E_{in}}\sqrt{m_{||}(\mathcal{E}_g^{\prime})^3}}
\label{TI}
\end{equation}

\noindent Here $\Delta n$ is the injected carrier density. $E_{in}$ is the THz electric field inside the bismuth film. $E_{in}$ is related to the THz electric field in vacuum, $E_v$, by a factor $\alpha$: $E_{in} = \alpha E_v$. At room temperature, $\alpha$ is estimated to be $\sim$ 0.25. $\mathcal{E}_g^{\prime}$ is the energy barrier and approximately equates $\mathcal{E}_g + \mathcal{E}_F$. $\mathcal{E}_F$ is the Fermi energy. $m_1$, $m_2$ and $m_3$ are determined by $m_i^{-1}=m_{ic}^{-1}+m_{iv}^{-1}$ ($i$ = 1, 2, 3), where 1, 2, and 3 refer to the binary, bisectrix, and trigonal axes, respectively. $m_{ic}$ and $m_{iv}$ are the effective masses of the bottom of conduction bands and the top of valance bands, respectively. $m_{||}$ is determined by $m_{||}^{-1}$ = cos$^2$$\theta$/$m_1$ + sin$^2$$\theta$/$m_2$. Here $\theta$ is the angle between THz polarization and the principal axis.

We take the numerical values of $m_1$, $m_2$, $m_3$, $\mathcal{E}_g$ and $\mathcal{E}_F$ from the ref 4, where $m_1^{-1}=(1.4\times10^{-3}m_0)^{-1}+(6.5\times10^{-3}m_0)^{-1}$, $m_2^{-1}=(0.29m_0)^{-1}+(1.36m_0)^{-1}$, $m_3^{-1}=(7\times10^{-3}m_0)^{-1}+(2.97\times10^{-2}m_0)^{-1}$, $\mathcal{E}_g$ = 15 meV, and $\mathcal{E}_F$ = 27 meV. Here $m_0$ is the mass of electron.\cite{Bismuth_parameters} It is important to note that the ionization rate is nearly independent of the crystallographic direction when taking into account all 3 L-points. For simplicity, we take $\theta$ = 0 to determine $m_{||}$. After inputting these parameters into Eq. \ref{TI}, we obtain the tunneling rate $dn/dt$ as a function of THz peak electric field as shown in Fig. \ref{Fig6_SI}(a) (blue curve). It clearly shows a threshold like behavior with increasing field. Unfortunately, the threshold field strength is about 80 kV/cm, two times smaller than the threshold field strength $\sim$ 200 kV/cm determined by our experiments. We noticed that the parameters we used are from the measurements of bismuth single crystals. However, we measured bismuth thin films. It is reasonable to expect that there exist some differences of these parameters in bismuth single crystals and thin films. To capture the threshold behavior of the phonon excitation, we rescale the parameters in the exponential part of Eq. \ref{TI} to match the threshold field strength 200 kV/cm revealed in the Fig. 3 of main text. The black curve in Fig. \ref{Fig6_SI}(a) represents the result after doing rescaling. Now we can see the threshold behavior at field strength $\sim$ 200 kV/cm is reasonably reproduced. We present temporal profiles of tunneling rate $dn/dt$ at various THz peak field strengths in Fig. \ref{Fig6_SI}(b). By integrating $dn/dt$ over time, we can directly calculate the temporal profile of injected charge density $\Delta n$. The resulting curves are shown in Fig. \ref{Fig6_SI}(c). One can see that both quantities are negligible below 150 kV/cm, indicting the threshold like behavior. At the same time, the temporal profile of injected charge density $\Delta n$ develops a rapid rise edge. We used a step function $(1+erf(2\sqrt{ln2}(t-t_0)/w))/2$ to fit the rise time. Here $t_0$ is the time zero. $erf$ is the error function. $w$ is the rise width. The fitting result is presented in Fig. \ref{Fig6_SI}(d). The rise time is found to be $\sim$ 200 fs, which is fast enough to drive the $A_{1g}$ mode in bismuth. To further demonstrate this, we performed Fourier transformation (FT) of this rise on $\Delta n$  (Fig. \ref{Fig6_SI}(c)) in the range of -1 to 0.8 ps where includes the rise edge. The result is displayed in Fig. \ref{Fig6_SI}(e). One can see that the FT amplitude of the rise contains remarkable spectral weight at the frequency of the $A_{1g}$ mode (2.92 THz), which could be the driving force of the $A_{1g}$ mode. As the field strength below 200 kV/cm, the spectral weight at the frequency of the $A_{1g}$ mode becomes negligible. 

\clearpage

\begin{figure}[t!]
\includegraphics[clip,width=6in]{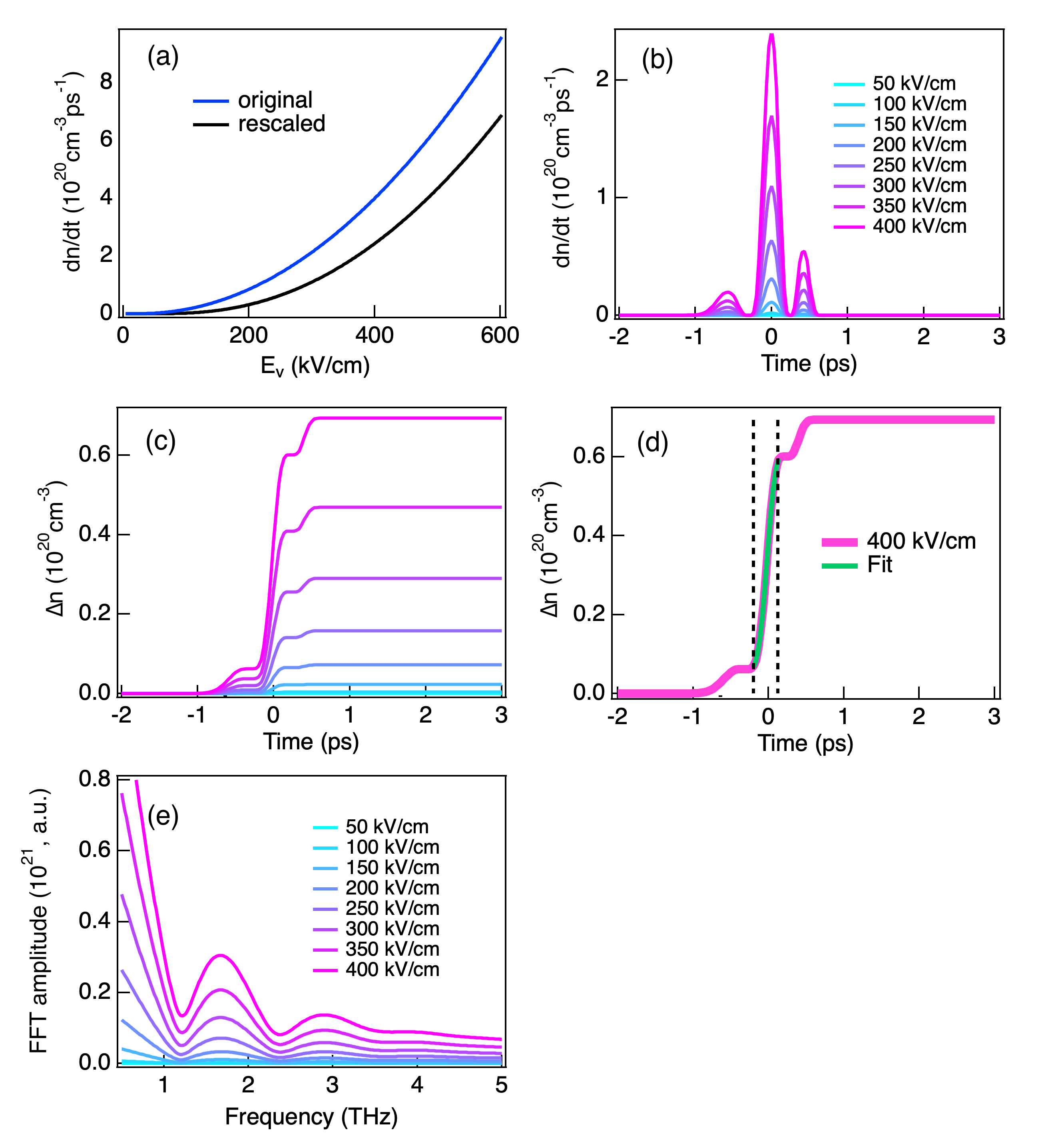}
\caption{(Color online) (a) Simulation of the tunneling rate $dn/dt$ as a function of THz electric field strength. (b) The temporal profile of the transient tunneling rate $dn/dt$ at various THz field strengths. (c) The temporal profile of the transient injected carrier density $\Delta n$ at at various THz field strengths. (d) The fit of the rise edge of the transient injected carrier density $\Delta n$ at a field strength of 400 kV/cm. (e) Fourier transformation amplitude of the rise on $\Delta n$. The Fourier transformation was performed on data shown in Fig. \ref{Fig6_SI}(c) in the range of -1 to 0.8 ps where includes the rise. }
\label{Fig6_SI}
\end{figure}

\clearpage

\section{S\lowercase{upporting} N\lowercase{ote} 7:  P\lowercase{honon dynamics simulated by tunnel ionization model} }

In Supporting Note 6, we obtained the temporal profile of transient injected carrier density $\Delta n$ via Eq. \ref{TI}. As discussed in the main text, the tunnel-ionization-induced sharp rise in charge density $\Delta n$ near time zero is sufficiently rapid to drive the $A_{1g}$ mode in bismuth. Consequently, we can use $\Delta n(t)$ from Fig. \ref{Fig6_SI}(c) to solve the following phonon motion equation \cite{DECP_1992}:

\begin{equation}
\frac{d^2Q}{dt^2}+\gamma\frac{dQ}{dt}+\omega^2_{0}Q=k\Delta n(t)
\label{anharmonic}
\end{equation}

\noindent Here, $Q$ is the driven phonon displacement. $\gamma$ is the damping coefficient. In our simulation, we set $\gamma$ = 0.47 THz THz to describe the phonon damping in real data of main text Fig. 1(b). The angular frequency of the $A_{1g}$ mode $\omega_0$ = 2$\pi$$\times$2.92 THz. The constant $k$ linearly relates the change in carrier density to the electrostatic force on the $A_{1g}$ mode, so the term $k\Delta n(t)$ will determine the strength of the phonon-driving force. The resulting simulation is shown in Fig. \ref{Fig7_SI}.

One can see our simulation captures several key ingredients of our experimental observations. First, the simulated phonon displacement $Q$ in Fig. \ref{Fig7_SI}(a) exhibits a rapid rise near time zero, which explains the rapid rise observed in THz-induced reflectivity change $\Delta R/R$. Second, the simulated phonon oscillations accurately reproduce the experimental observations. As illustrated in Fig. \ref{Fig7_SI}(b), the simulated oscillations closely match those seen in the experimental data. Third, as shown in Fig. \ref{Fig7_SI}(a), the simulated phonon displacement $Q$ exhibits clear threshold behavior. For THz fields above 200 kV/cm, $Q$ becomes significant and is comparable to the value at 400 kV/cm, while for fields below 200 kV/cm, $Q$ is negligible. To emphasize this threshold effect, we extracted the phonon amplitude from $Q$ at various THz field strengths using a cosine function cos$\omega_0t$. Fig. \ref{Fig7_SI}(c) plots the simulated phonon amplitude versus THz field strength, alongside the experimental data. The simulated amplitude shows a distinct threshold behavior that deviates from a simple $E^2$ dependence. Most importantly, our simulation closely follows the field dependence observed experimentally, further validating our tunnel model for describing phonon excitation in bismuth with THz pulses. Finally, we emphasize that our simulation does not account for carrier relaxation, which is essential for accurately capturing the extended carrier relaxation dynamics observed in $\Delta R/R$ at longer timescales. In this study, we focus specifically on the phonon excitation driven by the THz-induced charge tunneling process near time zero. Our simulation successfully reproduces the key features of THz pulse-driven Raman-active phonons in bismuth.

 \begin{figure}[hbt]
\includegraphics[clip,width=6in]{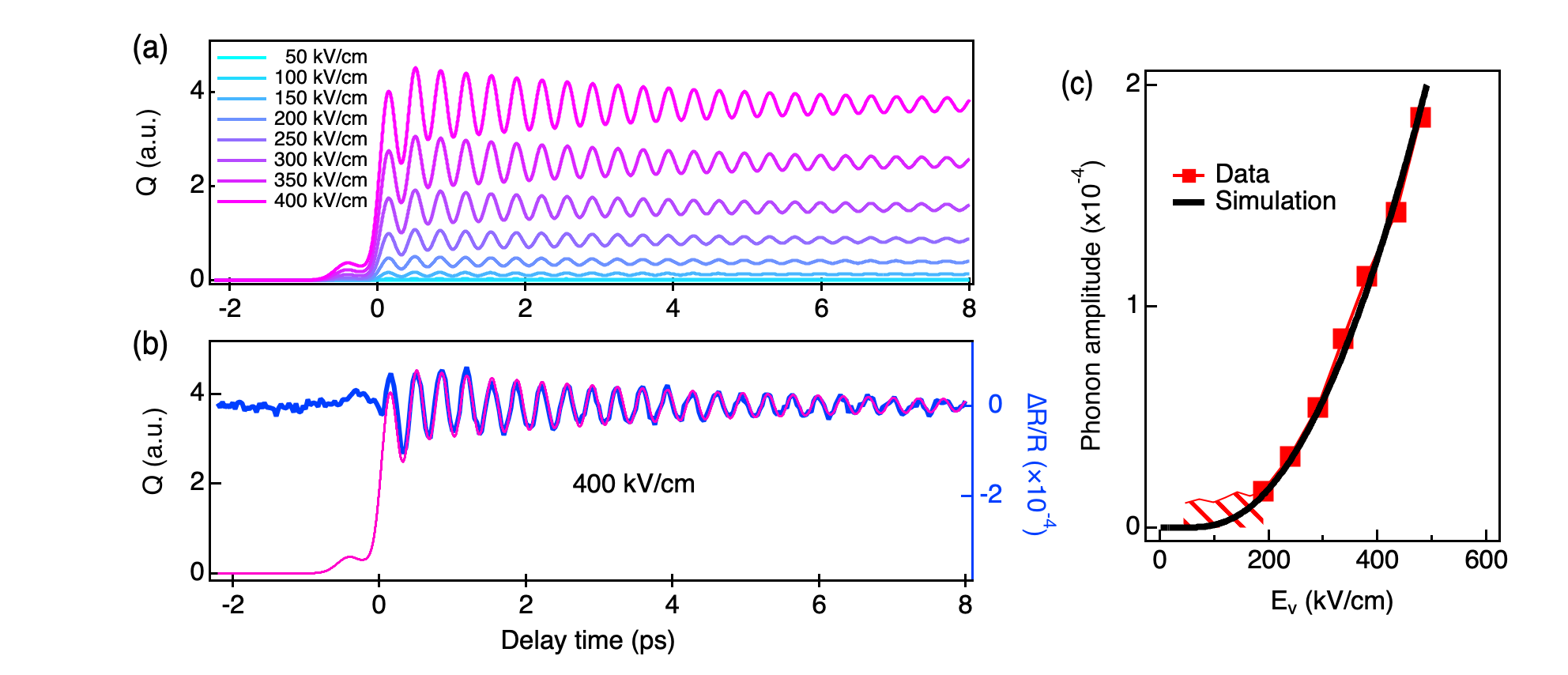}
\caption{(Color online) (a) Simulation of phonon dynamics under various THz field strengths with tunnel ionization model. (b) Phonon dynamics at 400 kV/cm. The experimental phonon oscillation at the same field strength is also plotted for comparison. (c) Simulated phonon amplitude as a function of THz field strength plotted together with the phonon amplitude determined by our experiments. }
\label{Fig7_SI}
\end{figure}

\clearpage

\section{S\lowercase{upporting} N\lowercase{ote} 8:  M\lowercase{ore evidence to exclude the} TH\lowercase{z sum frequency model} }

In this section, we provide more evidence to exclude two types of two-photon absorption process as the mechanism to drive the $A_{1g}$ mode in bismuth with THz pulse. First, let us examine if the two-photon absorption via charge excitation is responsible for phonon excitation. As bismuth is a semimetal (Fig. \ref{Fig8_SI_1}(a)), the onset of interband transitions is larger than $E_F + E_g$. If the two-photon absorption via charge excitation were responsible for carrier generation across the gap, the combined energy of the two photons would need to exceed $E_F + E_g$ ($\sim$ 10.1 THz). However, as shown in Fig. \ref{Fig8_SI_1}(b), the Fourier spectrum of $E^2(t)$ is far below the onsite energy scale of interband transitions in bismuth. Consequently, it is unlikely for this two-photon absorption process to generate carriers in the conduction bands and launch the phonon via the DECP mechanism.

Second, we can further look at if another two-photon absorption process---the THz sum frequency process---can explain our results. This is a Raman-type nonlinear photonics process. The phonon motion equation of THz sum generation model can be wrote as \cite{THzsum_theory111_2018}: 

\begin{equation}
\frac{d^2Q}{dt^2}+\gamma\frac{dQ}{dt}+\omega^2_{0}Q=\varepsilon_0RE^2(t)
\label{SumTHz}
\end{equation}

\noindent Here $Q$ is the driven phonon displacement. $\gamma$ is the damping coefficient. $\omega_0$ is the angular frequency of the driven phonon mode. $\varepsilon_0$ is the vacuum permittivity. $R$ is the Raman tensor at the driven phonon frequency. $E(t)$ is the electric field of THz pulse. Note that in this model, the driving force of the phonon is provided by $E^2(t)$, rather than $E(t)$. In Fig. \ref{Fig8_SI_1}(b), we show the Fourier transformation (FT) spectra of $E^2(t)$. One can see that, at the frequency of the $A_{1g}$ mode, there is nonzero spectral weight, which in principle could drive the mode. However, whether the phonon oscillations predicted by the THz sum-frequency model manifest in the experimental data depends entirely on whether the relative strength of the effective driving force, $\varepsilon_0RE^2(t)$, is sufficiently large and dominates over other driving mechanisms (such as $k\Delta n(t)$ discussed in Supporting Note 7).

 \begin{figure}[hbt]
\includegraphics[clip,width=5in]{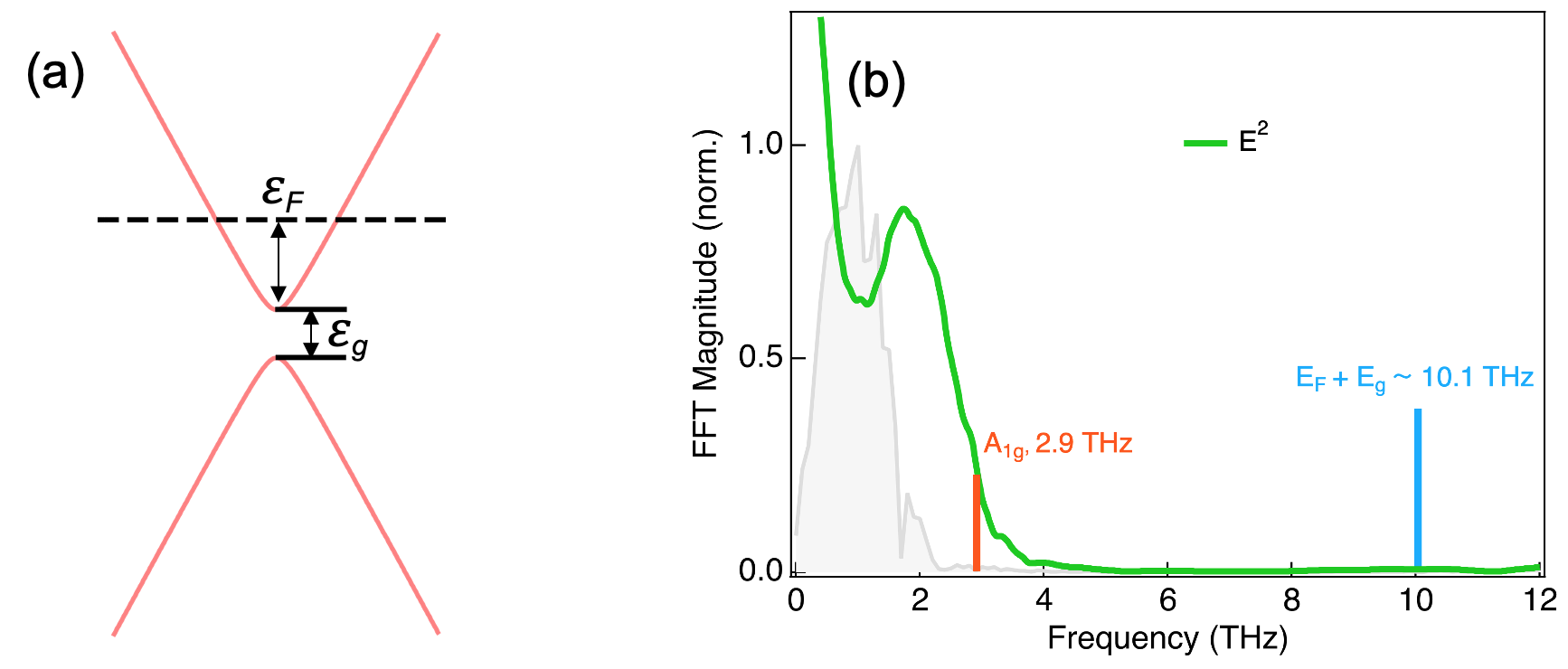}
\caption{(Color online) (a) Band structure schematic of bismuth near L points. (b) Fourier transformation magnitude of $E^2(t)$. Here $E(t)$ is the electric field waveform of THz pulse. The spectral magnitude of THz pulse $E(t)$ is indicated by the gray area. }
\label{Fig8_SI_1}
\end{figure}

\clearpage

We performed simulation of phonon dynamics using Eq. \ref{SumTHz}.  In our simulation, we set $\gamma$ = 0.47 THz, $\omega_0$ = 2$\pi$$\times$2.92 THz, and $E(t)$ as the THz pulse waveform shown in the main text figures. The simulated phonon dynamics as a function of delay time is shown in Fig. \ref{Fig8_SI_2}(a). One can see that in principle the spectral component at 2.9 THz in $E^2(t)$  can drive the $A_{1g}$ mode. However, our simulation based on the THz sum frequency model exhibits two critical discrepancies with our experimental observations. First, the phonon displacement $Q(t)$ (Fig. \ref{Fig8_SI_2}(a)) predicted by this model does not exhibit the rapid rise near time zero, which makes the rapid rise observed in $\Delta R/R$ unaccounted; in contrast, the tunnel ionization model (see Supporting Note 7) naturally produces a rapid initial rise in $Q(t)$, thereby accounting for the observed experimental behavior. Second, as illustrated in Fig. \ref{Fig8_SI_2}(a), the THz sum frequency model predicts a quadratic dependence of the driven phonon amplitude on the THz field strength, which is inconsistent with the experimentally determined phonon amplitude that reveal a threshold behavior and deviates from a simple $E^2$ dependence. Collectively, these discrepancies strongly suggest that the THz sum frequency model is unable to capture the main features of our experimental data. Combining the information from the simulations with both models, we conclude that the THz-driven tunnel ionization, rather than the THz sum-frequency process, is responsible for driving the $A_{1g}$ model in bismuth. 

 \begin{figure}[hbt]
\includegraphics[clip,width=5.5in]{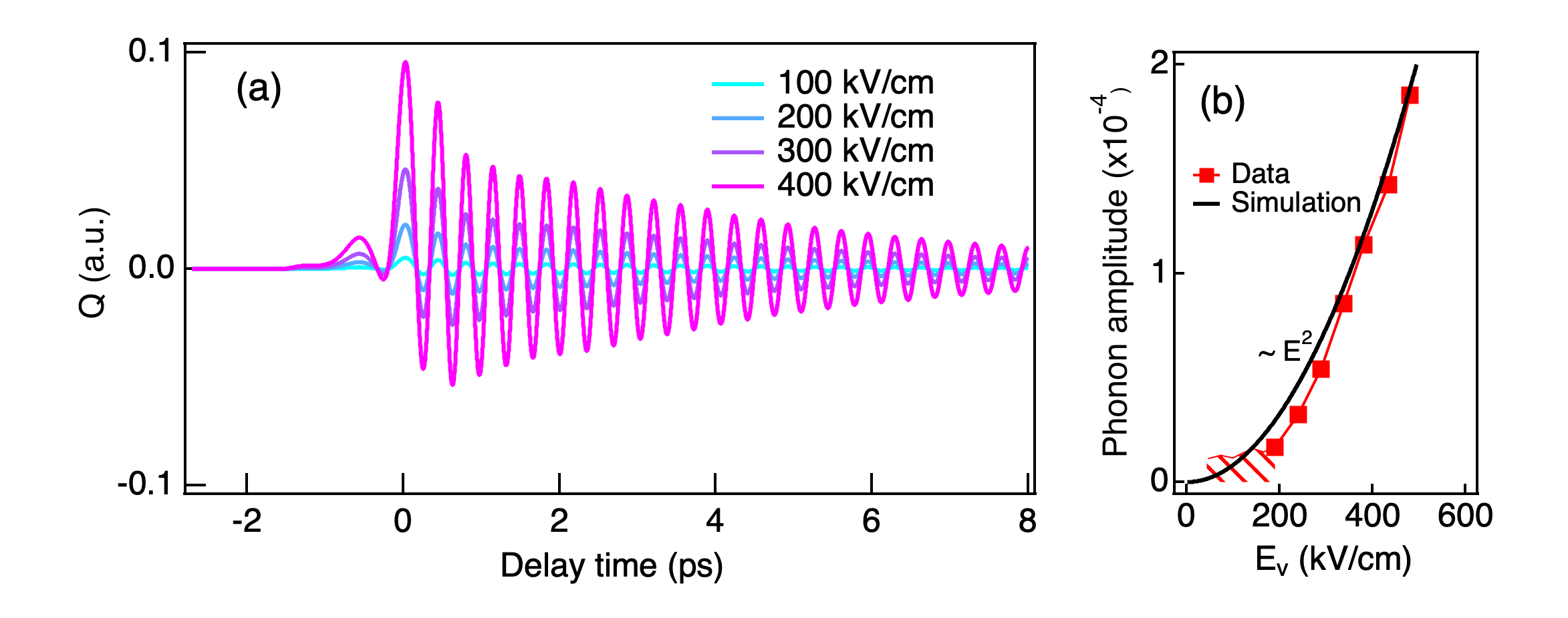}
\caption{(Color online) (a) Simulation of phonon dynamics under various THz field strengths with THz sum frequency model. (b) Simulation of driven phonon amplitude as a function of THz field strength plotted together with the phonon amplitude determined by our experiments.}
\label{Fig8_SI_2}
\end{figure}

\end{widetext}

\end{document}